# Direct observation of distinct bulk and edge nonequilibrium spin accumulation in ultrathin MoTe$_2$

Fangchu Chen[1,2], Kamal Das[3], Bowen Yang[1,2], Chuangtang Wang[4], Shazhou Zhong[1,2], Diana Golovanova[3], He Ren[1,5], Tianyang Wang[6,7], Xuan Luo[6], Yuping Sun[6,8,9], Liuyan Zhao[4], Guo-Xing Miao[1,5], Binghai Yan[3,10], and Adam W. Tsen[1,11]*

[1] Institute for Quantum Computing, University of Waterloo, Waterloo, ON, N2L 3G1, Canada

[2] Department of Physics and Astronomy, University of Waterloo, Waterloo, ON, N2L 3G1, Canada

[3] Department of Condensed Matter Physics, Weizmann Institute of Science, Rehovot 7610001, Israel,

[4] Department of Physics, University of Michigan, Ann Arbor, MI, USA

[5] Department of Electrical and Computer Engineering, University of Waterloo, Waterloo, ON, N2L 3G1, Canada

[6] Key Laboratory of Materials Physics, Institute of Solid State Physics, HFIPS, Chinese Academy of Sciences, Hefei, 230031, China

[7] University of Science and Technology of China, Hefei, 230026, China

[8] Anhui Province Key Laboratory of Low-Energy Quantum Materials and Devices, High Magnetic Field Laboratory, HFIPS, Chinese Academy of Sciences, Hefei, 230031, China

[9] Collaborative Innovation Center of Advanced Microstructures, Nanjing University, Nanjing, 210093, China

[10] Department of Physics, the Pennsylvania State University, University Park, PA 16802, USA

[11] Department of Chemistry, University of Waterloo, Waterloo, ON, N2L 3G1, Canada

*awtsen@uwaterloo.ca

## Abstract:

Low-symmetry two-dimensional (2D) topological materials such as MoTe$_2$ host efficient charge-to-spin conversion (CSC) mechanisms that can be harnessed for novel electronic and spintronic devices. However, the nature of the various CSC mechanisms and their correlation with underlying crystal symmetries remain unsettled. In this work, we use local spin-sensitive electrochemical potential measurements to directly probe the spatially dependent nonequilibrium spin accumulation in MoTe$_2$ flakes down to four atomic layers. We are able to clearly disentangle contributions originating from the spin Hall and Rashba-Edelstein effects and uncover an abundance of unconventional spin polarizations that develop uniquely in the sample bulk and edges with decreasing thickness. Using *ab-initio* calculations, we construct a unified understanding of all the observed CSC components in relation to the material dimensionality and stacking arrangement. Our findings not only illuminate previous CSC results on MoTe$_2$ but also have important ramifications for future devices that can exploit the local and layer-dependent spin properties of this 2D topological material.

## Introduction:

In nonmagnetic materials with strong spin-orbital coupling, the passage of charge current can generate pure spin currents via the spin Hall effect (SHE)[1–3] as well as nonequilibrium spin accumulation from both the SHE and Rashba-Edelstein effect (REE)[4–6]. These charge-to-spin conversion (CSC) mechanisms can be harnessed for the electrical generation and manipulation of spins in spintronic devices[7–10], such as in spin-orbit torque (SOT) magnetic memory and oscillators, while their inverse effects can be further used for electrical spin sensing[11,12]. In conventional CSC materials, the charge current, spin current, and spin polarization directions are constrained by the crystal symmetry to be mutually orthogonal[13,14], which limits the configuration of magnetic moments that can be created or detected. Recently, low-symmetry two-dimensional (2D) topological materials like $MoTe_2$ and $WTe_2$ have been reported to host unconventional (nonorthogonal) CSC, making them particularly appealing for next-generation spintronic applications[15–21]. At the same time, strong nonlinear anomalous Hall transport have been uncovered in these compounds[22–25], which results from the current-induced magnetization and can be directly exploited for high-frequency wave rectification[26–28]. There is thus diverse need to fully elucidate the underlying CSC mechanisms and their relationship with the crystal symmetries. While there have been important studies on the SOTs[15,18–20] and spin currents/polarizations in such systems[29–36], unresolved issues persist due to several challenges. First, as both the SHE and REE can yield the same torque and spin components, these effects can be extremely difficult to disentangle. Second, spin currents are hard to sense directly, and so their directions are often inferred from the device geometry. Third, it is unsettled what symmetries remain in the ultrathin limit and which/why certain unconventional CSC components are allowed.

In this work, we set out to directly detect the local current-induced spin accumulation (and associated in-plane spin polarization directions) in $MoTe_2$ flakes across a range of thicknesses. While magneto-optical techniques can be applied to directly image the spin accumulation in semiconductors[37], the signal is much weaker in metallic systems and is sensitive only to out-of-plane polarization[38]. We instead electrically probe the spin-dependent electrochemical potential[39,40] using a combination of ferromagnetic and nonmagnetic electrodes in contact at various positions on the sample and in-plane field control of the ferromagnetic spin orientation. Since the REE produces a uniform bulk magnetization while the SHE generates opposite spin polarizations at the sample boundary (perpendicular to the spin current), we can clearly distinguish between these two effects. By measuring from ~20-nm-thick flakes down to four atomic layers, we observe spin accumulation throughout the sample bulk from a conventional SHE and unconventional REE that persists for all thicknesses as well as distinct edge spin accumulation from unconventional SHEs that appear as thickness is lowered. By comparison with *ab-initio* calculations, the observed behavior can be explained by considering symmetry-breaking from a combination of reduced dimensionality and interlayer stacking shifts in thin flakes, the latter property being recently uncovered in cross-sectional electron microscopy[41]. Our study maps out the full thickness evolution of the spatially dependent nonequilibrium spin accumulation in $MoTe_2$ and elucidates the underlying crystal-symmetry-coupled CSC mechanisms that can be exploited for future electronic and spintronic applications.

The layered topological semimetals $MoTe_2$ and $WTe_2$ with octahedral coordination are both prime candidates for CSC and nonlinear transport phenomena, although we focus on the former in this study. Below ~250K, thick $MoTe_2$ exists in the noncentrosymmetric $T_d$ phase[42,43] (see Fig. 1a, left panel), which contains mirror and glide plane symmetries normal to the *a*-axis and *b*-axis, respectively, and a screw axis along the *c*-axis[24,31]. The corresponding $Pmn2_1$ space group allows for both a REE of the form: $S_i = \alpha_{ij} E_j$ with nonzero matrix elements $\alpha_{ab}$ and $\alpha_{ba}$, and a SHE of the form: $j_i^k = \sigma_{ij}^k E_j$ with nonzero spin Hall conductivity elements $\sigma_{bc}^a, \sigma_{ac}^b, \sigma_{ab}^c, \sigma_{cb}^a, \sigma_{ca}^b, \sigma_{ba}^c$ [44]. Here, $(E_a, E_b, E_c)$ is the applied electric field along the *a*, *b*, and *c* axes, respectively, $S_i$ is the induced polarization component, and $j_i^k$ is the spin current density with direction $i$ and spin polarization along $k$. As the charge current, spin current, and spin polarization directions are always mutually orthogonal, the thick crystal structure can only yield conventional CSC, in principle. For thin (or surface) layers, the glide plane and screw axis symmetries are nominally broken, and so additional unconventional elements may become available[13,31]. In reality, the structure of MoTe$_2$ flakes is more complex. Previous Raman studies have reported both the presence and absence of the $T_d$ phase at room temperature in few-layer samples[18,41,45,46], while a recent transmission electron microscopy study has shown substantial interlayer stacking rearrangements along the *b*-axis even for thicknesses of tens of nm [41], which would also break the glide plane and screw axis symmetries. The same microscopy study also may observe slight shifts along the *a*-axis, thus subtly breaking the last remaining mirror symmetry. The CSC mechanisms and components are hence intimately tied to the detailed microscopic structure.

## Results:

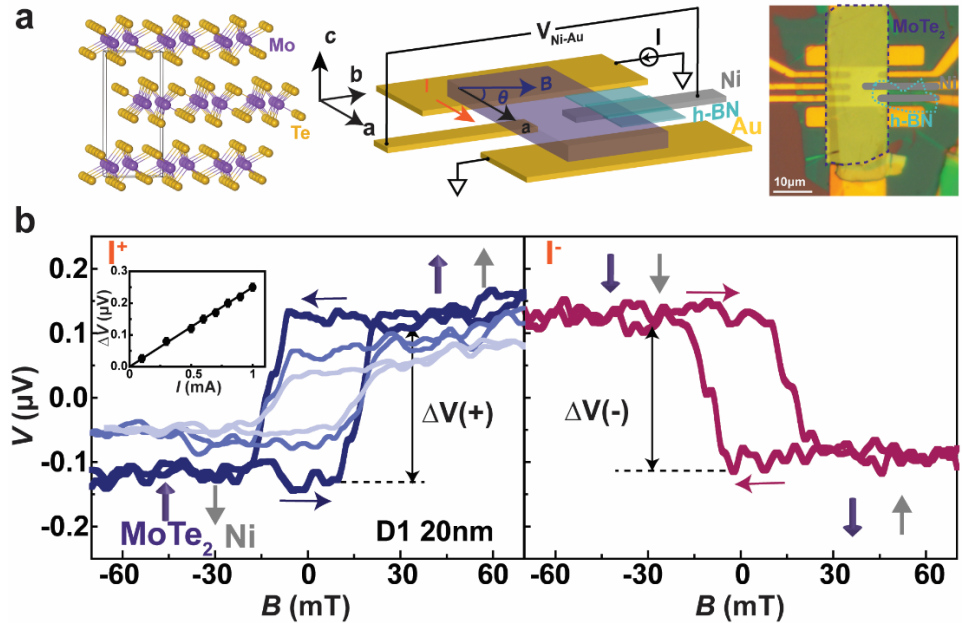

**Figure 1 | Concept and demonstration of spin accumulation measurements on MoTe$_2$ flakes. a**, Left panel: crystal structure of $T_d$-MoTe$_2$. Middle panel: schematic of the device and measurement setup. Purple and cyan flakes represent MoTe$_2$ and blocking h-BN, respectively. Gray and gold electrodes represent Ni and Au, respectively. The charge current is applied along the MoTe$_2$ *a*-axis and the magnetic field can be applied along any direction within the *ab*-plane to polarize the Ni. The voltage measured between the Ni and Au Hall leads senses the spin-dependent electrochemical potential locally in MoTe$_2$ at the Ni contact position. A difference is detected for current-induced spin polarization aligned or anti-aligned with the Ni polarization. Right panel: optical image of example MoTe$_2$ device with Ni contact to the MoTe$_2$ "bulk". The MoTe$_2$ flake and blocking h-BN are outlined by purple and cyan dashed line. The Ni electrode is colorized with gray for convenience. **b**, Representative spin accumulation results taken at 2K for Device 1 (20nm). Left panel: background-subtracted spin-dependent electrochemical potential $V$ as a function of magnetic field $B$ ($\theta = -22°$) for various positive currents ($I$ = 0.3, 0.6, 1mA). A "positive" hysteresis loop is seen due to the change of the magnetization state of the Ni relative to the local MoTe$_2$ spin polarization. The height of the hysteresis scales with current. The upper left inset shows the extracted hysteresis voltage $\Delta V$ (defined in the main panel) vs. current, indicating a linear relationship. Right panel: $V$ vs. $B$ for negative current flow ($I$ = -1mA) exhibits a "negative" hysteresis loop.

The middle panel of Fig. 1a shows the principle of our device geometry consisting of nonmagnetic Au current leads as well as ferromagnetic Ni and Au voltage leads in a Hall configuration and MoTe$_2$ crystals of varying thickness transferred on top (see Methods for fabrication details). We aligned our devices to pass current along the MoTe$_2$ *a*-axis, which was confirmed by magnetoresistance and/or optical second harmonic generation measurements (see Methods, Fig. S8, and Supplementary Section 1). The Ni serves as a local spin sensor, whose magnetization direction can be polarized in any direction within the *ab*-plane via a small magnetic field. By measuring the potential difference between the Ni and Au voltage leads, we can determine the spin-dependent electrochemical potential at the Ni position[39,40]. The Ni can either be contacted near the MoTe$_2$ edge to sense the edge spin accumulation or be extended deeper into the sample with the edge blocked by insulating hexagonal boron nitride (h-BN) to sense only the bulk spin accumulation at the electrode tip, as is the case shown in the schematic. A representative optical image for a bulk Ni contact device (Device 1, see Supplementary Table 1 for a list and description of all devices measured) is shown in the right panel of Fig. 1a.

Figure 1b demonstrates our measurement scheme on Device 1. As shown in the left panel, for positive current along the *a*-axis (defined by a common convention with respect to the electrode orientation), sweeping the magnetic field at 2K generally traces out hysteretic behavior for the Ni-Au voltage (see Fig. S9 for background subtraction procedure), which can be understood from the inset markings. The accumulated spin polarization in MoTe$_2$ is determined by the current direction, while the Ni polarization can be oriented and switched by the magnetic field. When the two are parallel (antiparallel), a positive (negative) voltage is detected after background subtraction[39]. The switching occurs for field values that are consistent with the coercive fields for the Ni film (see Fig. S10 and Methods). While the height of the hysteresis (difference between the two saturated voltage levels at positive and negative field) increases with the current magnitude, the sense of the hysteresis changes with the current direction (see Fig. 1b, right panel). The left inset of Fig. 1b further shows that the hysteretic voltage change Δ*V* follows a linear relationship with current, thus indicating our measurement is a direct probe of the linear-response spin accumulation in MoTe$_2$ at the Ni contact induced by current flow. We note that we have carefully ruled out the possibility of anomalous Hall effects in Ni as the origin of the hysteresis (see discussion in Supplementary Section 2). We have also performed the field sweeps at higher temperatures and find that Δ*V* decreases monotonically (see Fig. S11). This effect is qualitatively similar to that seen in previous potentiometric measurements of spin accumulation[40], and so we maintain 2K to maximize the signal in our experiment.

By selective placement of the Ni tip(s) at different locations on the sample, we can determine the local spin polarization and further distinguish between the REE and SHE mechanisms for nonequilibrium spin accumulation. Figure 2a shows a device schematic for Device 2 (19nm) where four Ni electrodes contact the left/right bulk and left/right edges of the sample (see Fig. S12 for optical image of device). There is a corresponding Au electrode for each Ni to allow for separate and balanced Hall voltage measurements. The field sweeps for each Ni contact with positive *a*-axis current flow and field orientation along the *b*-axis are plotted in the upper panels in Fig. 2b. We have normalized the voltage signal by current density for easy comparison across devices. For the left and right bulk contacts, we observe hysteresis loops of the same sense (labeled negative according to our definition in Fig. 1b), which indicates a sizeable spin population with polarization component along the negative *b*-axis ($S_b$) present in the sample interior. However, no hysteresis is observed for both edge contacts, thus indicating insignificant spin accumulation there within our measurement resolution.

We have performed similar measurements on Device 3 (23nm) (see Fig. S12 for optical image) for field sweeps along the *a*-axis and the results are shown in the lower panels in Fig. 2b. While the edges again show immeasurable spin accumulation, positive hysteresis loops are now observed for both bulk contacts, thus indicating the spins in the sample interior have a substantial spin component along the positive *a*-axis ($S_a$) as well. As this direction is parallel to the charge current flow, this already points to an unconventional

CSC mechanism. In Fig. 2c, we have plotted the height of the hysteresis for each measurement (normalized to current density) as a function of the position of the Ni contact along the $b$-axis relative to the sample center. For both $S_b$ and $S_a$, this value is similar for the opposing bulk contacts, suggesting that a relatively strong and symmetric spin accumulation develops in the interior, which then rapidly falls to zero at the flake boundaries.

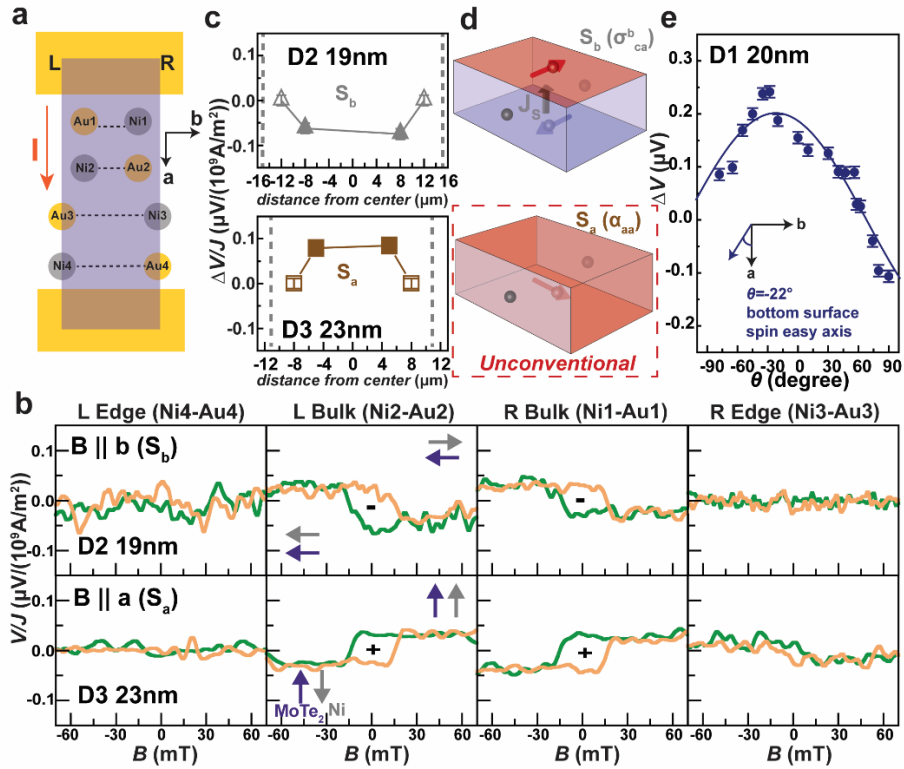

**Figure 2 | Position-dependent spin accumulation measurements of thicker MoTe₂. a**, Device schematic for local bulk and edge spin accumulation detection at left and right sides of a single sample. Separate Ni-Au electrode pairs allow for balanced Hall measurements. **b**, Top panels: spin accumulation measurements for each Ni contact on Device 2 (19nm) with $B \parallel b$ to detect for $b$-axis spin polarization ($S_b$). Negative hysteresis loops are observed for both bulk Ni contacts, but negligible hysteresis is seen for edges. Bottom panels: same for Device 3 (23nm) with $B \parallel a$ to detect for $a$-axis spin polarization ($S_a$). Positive hysteresis loops are observed for both bulk Ni contacts, but negligible hysteresis is seen for edges. Each voltage signal is normalized to current density $j$. **c**, Spin accumulation signal $\Delta V/j$ for each measurement in **b** as a function of the Ni contact position along the sample $b$-axis. Both $S_b$ and $S_a$ data appear relatively symmetric with respect to the sample center. The grey dashed lines indicate the position of the sample boundaries. **d**, Schematics of the two CSC mechanisms determined for thicker MoTe₂. The observed $S_b$ and $S_a$ spin polarizations in the sample bulk are assigned to conventional SHE $\sigma_{ca}^b$ and unconventional REE $\alpha_{aa}$, respectively, while the sample edges remain unpolarized. The unconventional CSC element is outlined by the red dashed lines. **e**, Magnetic field angle dependence of $\Delta V$ for bulk Ni contact taken on Device 1 (20nm). Solid line is fit to $\propto \cos(\theta + 22°)$, indicating the polarization axis for spins accumulated on the interior of the bottom surface.

Both the REE element $\alpha_{ba}$ and SHE element $\sigma_{ca}^b$ can give rise to the observed $b$-axis spin polarization (see Fig. S3). As discussed in Supplementary Section 3, symmetry dictates that the former would always exhibit the same spin polarization direction when probing the bottom surface of the sample (for fixed current and field orientations), while the latter can show opposite spin directions depending on how the flake is "flipped". Our consistent results obtained across many devices allow us to conclude that $\sigma_{ca}^b$ is responsible for $S_b$, which is a conventional SHE process allowed by the symmetry of the bulk crystal. Similarly, while both the REE element $\alpha_{aa}$ and SHE element $\sigma_{ca}^{a}$ can give rise to the unconventional a-axis spin polarization (see Fig. S3), we attribute $S_a$ to $\alpha_{aa}$ using the same symmetry arguments. Schematics of these two assigned processes are shown in Fig. 2d for the sample interior (where we place our bulk Ni contacts), while the edges remain unpolarized. Broken translational symmetry at the sample boundaries can possibly reduce the local $\sigma_{ca}^{b}$ and $\alpha_{aa}$ contributions, leading to undetectable edge spin accumulation.

When both $S_a$ and $S_b$ are finite, the spin axis within the *ab*-plane can also be directly measured. In Fig. 2e, we show the hysteretic voltage change $\Delta V$ for a bulk Ni contact as a function of the applied magnetic field angle $\theta$ relative to the *a*-axis for Device 1 (see Fig. S13 for complete field sweeps at various angles). We have fit the data to a cosine dependence and the maximum $\Delta V$ is extracted for $\theta = -22°$, which indicates the orientation of the spins on the interior of the bottom sample surface. We then expect that the corresponding spins polarization on the top surface to be pointed along $\theta = 22°$. Note that the spin axis varies across different samples (see Supplementary Table 2).

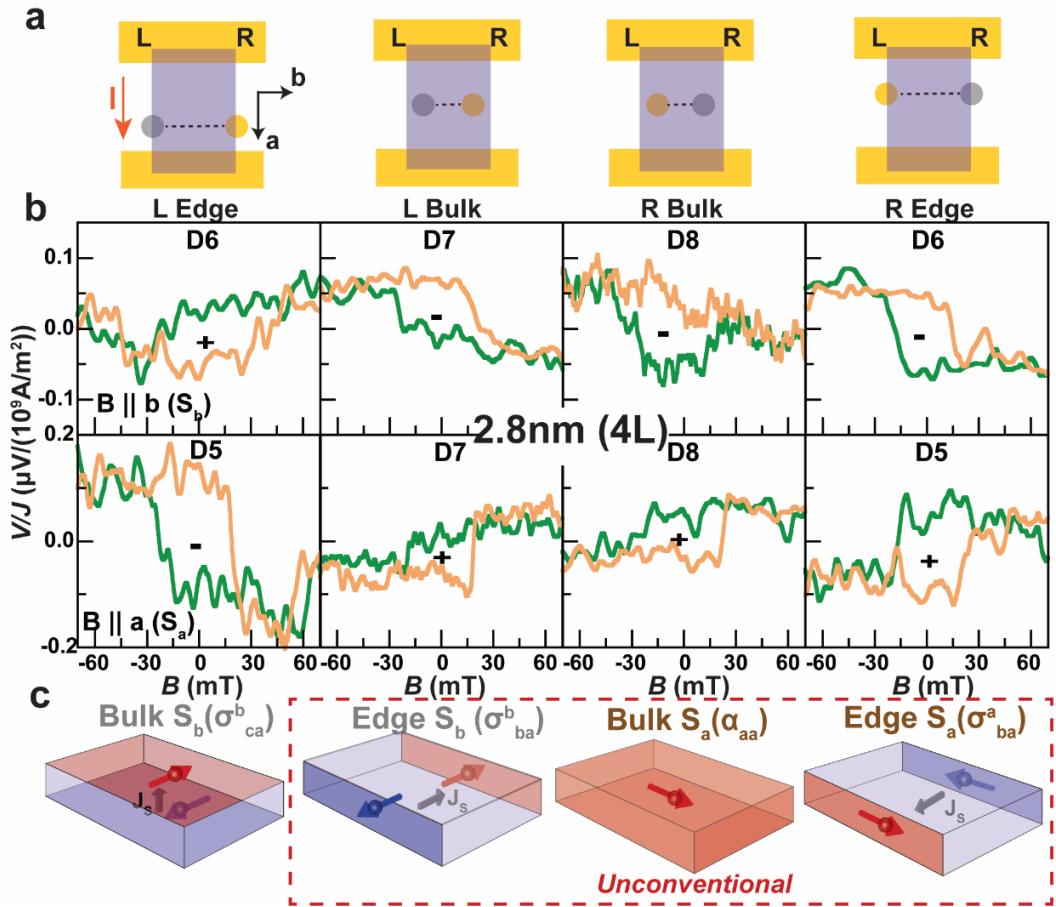

**Figure 3 | Distinct bulk and edge spin polarization in 2.8nm (four-layer) MoTe$_2$. a,** Device schematics for left edge, left bulk, right bulk, and right edge Ni contacts on 2.8nm MoTe$_2$ samples. Gold and gray circles indicate Au and Ni contacts, respectively. **b,** Spin accumulation measurements normalized to current density for each Ni contact with $B \parallel b$ to detect for $S_b$ (top panels) and $B \parallel a$ to detect for $S_a$ (bottom panels). The left and right bulk measurements are consistent with those seen for thicker flakes; however, left and right edges now exhibit finite and opposite hysteresis in both $S_b$ and $S_a$ configurations. **c,** Schematics of the four CSC mechanisms determined for 2.8nm MoTe$_2$. As in thicker flakes, SHE $\sigma_{ca}^{b}$ and REE $\alpha_{aa}$ account for the bulk spin polarizations $S_b$ and $S_a$, respectively, while additional $S_b$ and $S_a$ edge polarizations are attributed to SHE elements $\sigma_{ba}^{b}$ and $\sigma_{ba}^{a}$, respectively. The unconventional CSC elements are grouped and outlined by the red dashed lines.

The observation of unconventional spin components as well as contrasting bulk and edge effects in even relatively thick samples motivates further measurements down to lower thicknesses. We have probed the local spin accumulation in MoTe$_2$ down to 2.8nm, or four layers, the lowest thickness (highest sample resistivity) for which the hysteresis loops can still be clearly distinguished above the noise level by our technique. Due to the smaller areas of 2.8nm flakes, we have fabricated separate devices for bulk and edge Ni contacts (Devices 5-8). Figure 3a shows device schematics for various such contact configurations, while optical images are shown in Fig. S12. For each Ni position, the *b*- and *a*-axis field sweeps (probing for $S_b$ and $S_a$, respectively) are plotted in Fig. 3b. While the left and right bulk contacts both show negative hysteresis like that for the thicker sample (Device 2), the two edges now exhibit finite and opposite hysteresis—positive (negative) hysteresis is seen for the left (right) edge. For *a*-axis field sweeps, positive hysteresis is observed for both left and right bulk contacts (similar to Device 3), but again the edges show opposite hysteresis loops—negative (positive) for the left (right) edge. These results indicate that ultrathin layers not only inherit the CSC elements $\sigma_{ca}^b$ and $\alpha_{aa}$ from their thicker counterparts that give rise to the same spin polarization in the sample interior but further possess additional contributions that yield finite spin accumulation at the edges. When the edge spin polarizations are opposite, it indicates a SHE with spin current along the *b*-axis is the underlying mechanism. We thus attribute the edge $S_b$ and $S_a$ spin components in ultrathin MoTe$_2$ to the unconventional SHE elements $\sigma_{ba}^b$ and $\sigma_{ba}^a$. In contrast to $\sigma_{ca}^b$ and $\alpha_{aa}$, for these two processes the spins accumulated at the edges originate from separation that occurs throughout the width of the sample (and not just locally near the edges themselves). We thus expect the accumulation to be less sensitive to the precise atomic structure at the flake boundaries. All the CSC processes detected from 2.8nm MoTe$_2$ are illustrated in Fig. 3c.

## Discussion:

Our remaining discussion will be focused on a unified understanding of all the experimentally observed SHE and REE components with changing MoTe$_2$ thickness. To show this dependence explicitly, we have plotted the all the spin accumulation $\Delta V$ (for $S_b$ and $S_a$, bulk and edge) normalized to the current density as a function of thickness in Fig. 4a. When a finite signal is observed across all thicknesses, as in the case of the bulk measurements, the values decrease with increasing thickness, which can be understood from the expected scaling with sample resistivity (see discussion in Supplementary Section 4). Although the *b*-axis glide mirror plane $\overline{M_b}$ and *c*-screw-axis $\vec{S_c}$ symmetries present in thick MoTe$_2$ are nominally broken in thin samples, a recent transmission electron microscopy study has reported changes in the interlayer stacking along the *b*-axis for MoTe$_2$ flakes with tens of nm thickness[41], the same thickness range where we observe the unconventional component $\alpha_{aa}$ in our experiment. This leads to a mixed phase between the pure 1*T'* or $T_d$ stacking configurations, thereby breaking both $\overline{M_b}$ and $\vec{S_c}$ symmetries explicitly. The same microscopy study also may have observed slight shifts along the *a*-axis. Although the magnitude of the *a*-axis shifts is much smaller than those for the *b*-axis, it would lead to subtle breaking of the last remaining mirror plane $M_a$ symmetry as well.

We have modeled the effects of reduced thickness and interlayer shifts using *ab-initio* density functional theory (DFT) calculations on pristine thick and 2.8nm MoTe$_2$ with various stacking configurations that reflect the microscopy results of ref 41 (see Methods and Supplementary Section 5). In the left column of Fig. 4b, we consider a particular 2.8nm structure with ideal $T_d$ stacking. This is captured by the alternating shifts of the Te atoms shown in the inset above. In the panels below, we have plotted all the relevant SHE and REE tensor components as function of the Fermi level calculated for this structure in solid lines. For comparison, we also show the corresponding plots for a thick crystal in the pure $T_d$ phase using dashed lines.

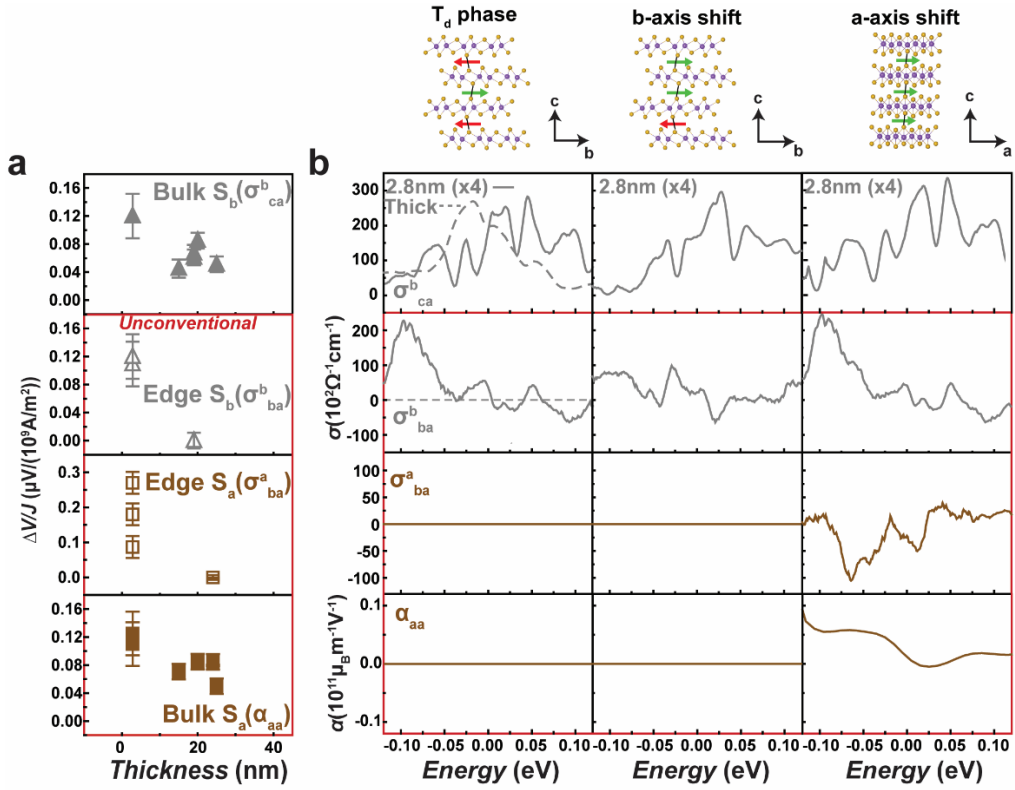

**Figure 4 | Thickness dependence of various spin polarization signals and DFT calculations of the CSC components. a**, Bulk and edge, $S_b$ and $S_a$ spin polarization signals normalized to current density vs. MoTe$_2$ thickness. The responsible CSC element for each is indicated in parentheses. **b**, Left panel: the relevant SHE and REE components vs. Fermi energy calculated for pristine 2.8nm and thick $T_d$-MoTe$_2$. Middle panel: same calculations for a mixed-phase 2.8nm MoTe$_2$ with layer shift along *b*-axis. Right panel: same calculations for a 2.8nm MoTe$_2$ with layer shift along *a*-axis. Schematics of the structures are shown in the insets above. The unconventional CSC elements are outlined in red in both **a** and **b**.

We see that reducing thickness alone does not lead to qualitative differences in the conventional component $\sigma_{ca}^b$ that gives rise to bulk $S_b$, as the relevant symmetries are already broken for the thick crystal structure. On the other hand, the unconventional component $\sigma_{ba}^b$ giving rise to edge $S_b$ is allowed for the 2.8nm system only, while $\sigma_{ba}^a$ and $\alpha_{aa}$ are forbidden in both structures. These results indicate that the appearance of edge $S_b$ in our 2.8nm devices can already be explained even without the inclusion of interlayer stacking changes, yet bulk and edge $S_a$ seen in 2.8nm MoTe$_2$ cannot be accounted for from the pristine structure.

To include the effects of the interlayer stacking, we have repeated the calculations in the middle panels of Fig. 4b for a particular low-energy mixed-phase $1T'$ / $T_d$ structure with *b*-axis shifts. Calculations performed for related structures as well as pure $1T'$ four-layer are shown Fig. S6. In the right panels of Fig. 4b, we show the calculated components for a plausible structure with *a*-axis shifts that gives the stack a slight tilt starting from the pristine $T_d$ four-layer. Taken together, these results indicate that the effects of stacking changes are relatively small for elements that are already nonzero in the pristine four-layer ($\sigma_{ca}^b$ and $\sigma_{ba}^b$). However, components $\sigma_{ba}^a$ and $\alpha_{aa}$ that give rise to edge and bulk $S_a$, respectively, can only be explained with the inclusion of *a*-axis shifts that break the last remaining $M_a$ symmetry. We have additionally tried other structures with *a*-axis shifts and find that $\sigma_{ba}^a$ and $\alpha_{aa}$ will generally be nonzero independent of the details of the shift since $M_a$ is always broken. This analysis thus captures the totality of our experimental results from a microscopic point of view.

Our direct measurements of spatially and thickness dependent nonequilibrium spin accumulation in MoTe$_2$ and supporting theoretical calculations put the CSC mechanisms on firm footing. While various current-induced spin polarizations have been previously deduced in this material using the inverse spin Hall effect[29–32,36], the indirect nature of these measurements together with multiple charge and/or spin current directions possible at the contacts and interfaces make definitive assignment of the SHE or REE components difficult. Our results also shed new light on earlier SOT experiments in MoTe$_2$. Specifically, it has been shown that *b*-axis spin polarization (for charge current along the *a*-axis) leads to efficient switching of nearby ferromagnets with in-plane anisotropy [20]; however, it was unclear whether this was generated by the SHE component $\sigma_{ca}^b$ or REE component $\alpha_{ba}$. This work indicates that $\sigma_{ca}^b$ is the predominant contributor. Additionally, SOT ferromagnetic resonance measurements have possibly detected small torque components generated by *a*-axis spin polarization[18]. If these torques are indeed finite, it most likely originates from $\alpha_{aa}$ caused by $M_a$-symmetry-breaking. Finally, an extremely large *c*-axis nonlinear anomalous Hall effect has been reported in both thin and thick MoTe$_2$[24]. As the anomalous Hall voltage develops as a consequence of the current-induced magnetization, it is also consistent with our direct observation of REE $\alpha_{aa}$ in even relatively thick samples.

By use of a local spin accumulation sensor, we have uncovered a plethora of unconventional spin polarizations in the low-symmetry topological material MoTe$_2$ that develop nonuniformly across the sample and can be tuned with thickness. Our work paves the way for the deliberate design of spatially dependent SOT or inverse SHE/REE device geometries that can fully take advantage of the distinct bulk and edge spin accumulation and/or thickness-dependent CSC processes in 2D MoTe$_2$ to realize more exotic functionalities in the future.

## Methods:

**Crystal Synthesis**
1T'-MoTe$_2$ single crystals were grown by the flux method using Te as a solvent. Mo (Alfa Aesar, 99.9%), Te (Alfa Aesar, 99.99%) powders were ground and placed into alumina crucibles in a ratio of 1:25 and sealed in a quartz ampoule. After the quartz ampoule was heated to 1050°C and held for 2 days, the ampoule was slowly cooled to 900°C and centrifuged. Shiny and plate-like crystals with lateral dimensions up to several millimeters were obtained.

**Device Fabrication**
Au (35nm)/Ti (5nm) and Ni (35nm)/Ti (5nm) electrodes in a Hall-bar geometry were pre-patterned on Si wafers with 285-nm-thick SiO$_2$ using conventional photolithography and electron-beam deposition. h-BN (HQ graphene) and MoTe$_2$ flakes were exfoliated onto blank SiO$_2$/Si wafers. The thickness of the MoTe$_2$ was determined by optical reflection contrast and atomic force microscopy (see Fig. S14). We have maintained the orientation of the bulk crystal during the exfoliation process and aligned the rectangular MoTe$_2$ flakes with the electrode pattern. This allows for current injection along *a*-axis, which can be confirmed by measuring the magnetoresistance or optical second harmonic generation, as we have done in our previous work[24]. After the desired flakes were identified, a polymer stamp coated with polycarbonate was used to sequentially pick up the entire h-BN/ MoTe$_2$/ h-BN heterostructure to avoid contamination between the layers. Prior to heterostructure transfer, the chip with electrodes was immersed in ammonia for five minutes to remove excess surface nickel oxide and then ultrasonicated in acetone and isopropanol. The heterostructure was then aligned and transferred onto the pre-patterned electrodes. The exfoliation, pick-up, and transfer processes were performed within a nitrogen-filled glovebox to avoid degradation of MoTe$_2$ in air.

**Optical Second Harmonic Generation Measurements**

The rotational anisotropy second harmonic generation (RA-SHG) measurements were performed to determine the crystal axes of the MoTe$_2$. The samples were cooled to 80K inside an optical cryostat. A 1200nm pulsed laser with a repetition rate of 200kHz was focused onto the MoTe$_2$-containing devices with a 20× objective lens at normal incidence. The reflected 600nm light was collected by the same objective lens. To isolate the SHG signal, the 1200nm fundamental light was filtered out using optical shortpass filters, allowing only 600nm SHG light to be detected by an electron-multiplying charge-coupled device. The incident light polarization was rotated using a half-wave plate, and the reflected light was selected to keep either parallel or crossed relative to the incident light polarization. The RA-SHG results are provided in Supplementary Section 1.

**Magnetotransport Measurements**

The spin accumulation measurements were carried out in a superconducting magnet Helium-4 cryostat. The devices were mounted on a rotator stage. An AC current with frequency 17Hz was passed along the *a*-axis of the MoTe$_2$ flake, and AC voltages were measured using an SR860/865 lock-in amplifier. For thicker samples (~20nm), we generally applied a current of 1mA, and the current density ranges between 1.5-2.0×10$^9$A/m$^2$ depending on the flake geometry. For four-layer samples, we applied a current between 0.1-0.2mA, and the current density ranges between 3-8×10$^9$A/m$^2$. Except for the temperature-dependent measurements presented in Fig. S11, all measurements were performed at 2K.

**Magnetization Measurements**

The magnetic properties of the Ni film were measured using a vibrating sample magnetometer (Lake Shore 8607). A DC magnetic field was applied by an electromagnet. The sample was attached to a sample rod made of fluorocarbon-based polymer using GE Varnish. The background signal was removed by performing the same measurement with the bare rod and GE Varnish.

**Ab-initio Calculations**

We performed *ab-initio* density functional theory (DFT) calculation to estimate the spin Hall conductivity and Rashba-Edelstein coefficients. The DFT calculation was performed with the Vienna ab-initio simulation package (VASP) with a projector augmented wave (PAW) potential[47,48]. The exchange-correlation function was considered in the generalized gradient approximation (GGA) level with Perdew-Burke-Ernzerhof (PBE) functional[49]. The cut-off energy for the plane wave expansion is considered 350 eV. The lattice constants for the thick samples were taken to be $a = 3.477$ Å, $b = 6.335$ Å, and $c = 13.883$ Å[43]. The thin films have been realized as 4-layer slabs. For the thick samples and slabs, $12 \times 8 \times 4$ and $12 \times 8 \times 1$ k-grid was considered, respectively. To calculate different physical properties, a tight-binding model was constructed using the maximally localized Wannier functions using the VASP2WANNIER90 codes[50], which was further symmetrized using the WannierTools [42]. For the Wannier fitting, we considered the *p* orbitals of Te and *d* orbitals of Mo atoms. The spin Hall conductivities and the Rashba-Edelstein coefficients were calculated using in-house codes. The details of the calculations are provided in Supplementary Section 5.

# Acknowledgements:

We thank Prof. Ding Fu Shao at the Institute of Solid State Physics, Chinese Academy of Sciences for helpful discussions of our manuscript. A.W.T acknowledges support from the Dorothy Killam Fellowship and US Air Force Office of Scientific Research (FA9550-24-1-0360). A.W.T and G-X.M acknowledge support from the Natural Sciences and Engineering Research Council of Canada (NSERC) and the Transformative Quantum Technologies Program. B.Y. acknowledges the financial support by the Israel Science Foundation (ISF: 2932/21) and National Science Foundation through the Penn State Materials Research Science and Engineering Center (MRSEC) DMR 2011839. L.Z. acknowledges the support from the U.S. Department of Energy (DOE), Office of Science, Basic Energy Science (BES), under award No. DE-SC0024145. X.L. and Y.P.S. thank the support from the National Key R&D Program (Grant Nos. 2023YFA1607402, 2021YFA1600201), the National Natural Science Foundation of China (Grant Nos. U2032215, No. U1932217, and 12274412), and Systematic Fundamental Research Program Leveraging Major Scientific and Technological Infrastructure, Chinese Academy of Sciences under contract No. JZHKYPT-2021-08. K.D. acknowledges the financial support from Weizmann Institute of Science, Dean of Faculty fellowship and the Koshland Foundation.

## Author Contributions

F.C. and A.W.T. conceived and initiated the study. F.C. fabricated the MoTe$_2$ devices for transport, optical second harmonic generation, and magnetization measurements. F.C. performed the local, spin-sensitive electrochemical potential measurements with the assistance of B.Y and S.Z. H.R. and G-X.M. conducted the magnetization measurements. C.W. and L.Z. performed the optical second harmonic generation measurements. K.D., D.G., and B.H.Y. performed the *ab-initio* calculations. F.C., T.W., X.L., and Y.P.S. grew the 1T′- MoTe$_2$ crystals. F.C. and A.W.T. wrote the manuscript with the input of all authors.

## Competing Interests

The authors declare no competing interests.

## Data Availability Statement:

The authors declare that the data supporting the findings of this study are available within the paper and/or the supplementary information, and available from the corresponding author on reasonable request.

# Direct observation of distinct bulk and edge nonequilibrium spin accumulation in ultrathin MoTe$_2$

Fangchu Chen[1,2], Kamal Das[3], Bowen Yang[1,2], Chuangtang Wang[4], Shazhou Zhong[1,2], Diana Golovanova[3], He Ren[1,5], Tianyang Wang[6,7], Xuan Luo[6], Yuping Sun[6,8,9], Liuyan Zhao[4], Guo-Xing Miao[1,5], Binghai Yan[3,10], and Adam W. Tsen[1,11*]

[1] Institute for Quantum Computing, University of Waterloo, Waterloo, ON, N2L 3G1, Canada

[2] Department of Physics and Astronomy, University of Waterloo, Waterloo, ON, N2L 3G1, Canada

[3] Department of Condensed Matter Physics, Weizmann Institute of Science, Rehovot 7610001, Israel,

[4] Department of Physics, University of Michigan, Ann Arbor, MI, USA

[5] Department of Electrical and Computer Engineering, University of Waterloo, Waterloo, ON, N2L 3G1, Canada

[6] Key Laboratory of Materials Physics, Institute of Solid State Physics, HFIPS, Chinese Academy of Sciences, Hefei, 230031, China

[7] University of Science and Technology of China, Hefei, 230026, China

[8] Anhui Province Key Laboratory of Low-Energy Quantum Materials and Devices, High Magnetic Field Laboratory, HFIPS, Chinese Academy of Sciences, Hefei, 230031, China

[9] Collaborative Innovation Center of Advanced Microstructures, Nanjing University, Nanjing, 210093, China

[10] Department of Physics, the Pennsylvania State University, University Park, PA 16802, USA

[11] Department of Chemistry, University of Waterloo, Waterloo, ON, N2L 3G1, Canada

*awtsen@uwaterloo.ca

# Supplementary Section 1: Second harmonic generation identification of crystal axis

The point group *m* with electric-dipole SHG radiation is used to describe the nonlinear optical process for $T_d$-MoTe$_2$ (Nature Communications 12, 2049 (2021), ref 24). The nonlinear susceptibility tensor of electric-dipole tensor $\chi^{ED}_{ijk}$ is:

$$\chi^{ED}_{ijk} = \begin{pmatrix} \begin{pmatrix} 0 \\ \chi^{ED}_{xyx} \\ \chi^{ED}_{xzx} \end{pmatrix} & \begin{pmatrix} \chi^{ED}_{xyx} \\ 0 \\ 0 \end{pmatrix} & \begin{pmatrix} \chi^{ED}_{xzx} \\ 0 \\ 0 \end{pmatrix} \\ \begin{pmatrix} \chi^{ED}_{yxx} \\ 0 \\ 0 \end{pmatrix} & \begin{pmatrix} 0 \\ \chi^{ED}_{yyy} \\ \chi^{ED}_{yzy} \end{pmatrix} & \begin{pmatrix} 0 \\ \chi^{ED}_{yzy} \\ \chi^{ED}_{yzz} \end{pmatrix} \\ \begin{pmatrix} \chi^{ED}_{zxx} \\ 0 \\ 0 \end{pmatrix} & \begin{pmatrix} 0 \\ \chi^{ED}_{zyy} \\ \chi^{ED}_{zzy} \end{pmatrix} & \begin{pmatrix} 0 \\ \chi^{ED}_{zzy} \\ \chi^{ED}_{zzz} \end{pmatrix} \end{pmatrix}$$

In the normal incidence geometry, the resulting forms for RA-SHG for the parallel and crossed channels are:

$$I^{2\omega}_{\text{parallel}}(\phi) = \sin^2(\phi)\left(\chi^{ED}_{yyy}\sin^2(\phi) + (2\chi^{ED}_{xyx} + \chi^{ED}_{yxx})\cos^2(\phi)\right)^2,$$

$$I^{2\omega}_{\text{crossed}}(\phi) = \cos^2(\phi)\left(\chi^{ED}_{yxx}\cos^2(\phi) + (\chi^{ED}_{yyy} - 2\chi^{ED}_{xyx})\sin^2(\phi)\right)^2$$

where $I^{2\omega}_{\text{parallel}}(\phi)$ and $I^{2\omega}_{\text{crossed}}(\phi)$ are the SHG intensity for the parallel and crossed channels, respectively, and $\phi$ is the azimuthal angle relative to the *a*-axis of MoTe$_2$. These two forms are used to fit the experimental data. The crossed channel results for 2.8nm, 19nm, and 23nm MoTe$_2$ devices are presented in Fig. S1. These results exhibit distinct patterns, yet all can be fitted within the *m* point group with varying strengths of the tensor elements. Importantly, SHG is strictly prohibited with zero intensity when the SHG polarization is parallel to the *a* axis, which is normal to the mirror direction. Therefore, we can identify the *a* and *b* axes of $T_d$-MoTe$_2$ as indicated in the figure.

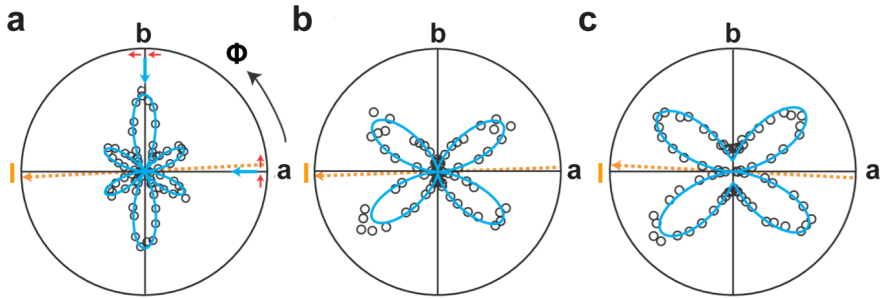

**Fig. S1**. RA-SHG patterns in the crossed channel for three MoTe$_2$ devices containing MoTe$_2$ of different thickness. Experimental patterns for **a** D6 (2.8nm), **b** D2 (19nm), and **c** D3 (23nm) MoTe$_2$ devices and their corresponding fittings are shown as circles and red lines, respectively. The blue and red arrows indicate the polarization of SHG (600nm) and incident (1200nm) light, respectively. The dashed orange line indicates the current flow direction, which is closely aligned with the MoTe$_2$ *a*-axis.

## Supplementary Section 2: Exclusion of the anomalous Hall effect

In order to exclude any anomalous Hall effect (AHE) due to possible current-shunting through the Ni contact as the origin of our spin accumulation signal, we have prepared and measured devices in two different geometries (see schematics in the upper panels of Fig. S2a and S2b below and optical images in Fig. S12). In both cases, the Ni is contacting the left bulk of $MoTe_2$. However, the Ni and Au electrodes extend from opposite directions in the two geometries. If there is current shunting at the Ni contact area and the AHE were the dominant mechanism, then the anomalous Hall current will flow in the same direction in both devices for a given Ni polarization. Measuring the voltage of the Ni relative to the Au will then trace out hysteresis loops of opposite sign when the field is swept (see schematics in the lower panels of Fig. S2a and S2b).

In contrast, our experimental data shows hysteresis loops of the same sign for both device geometries in either $B \parallel a$ or $B \parallel b$ field configuration (see Fig. S2c and S2d). This rules out the possibility of the AHE in Ni as the origin of our signal and corroborates that the observed hysteresis is indeed due to spin accumulation in $MoTe_2$.

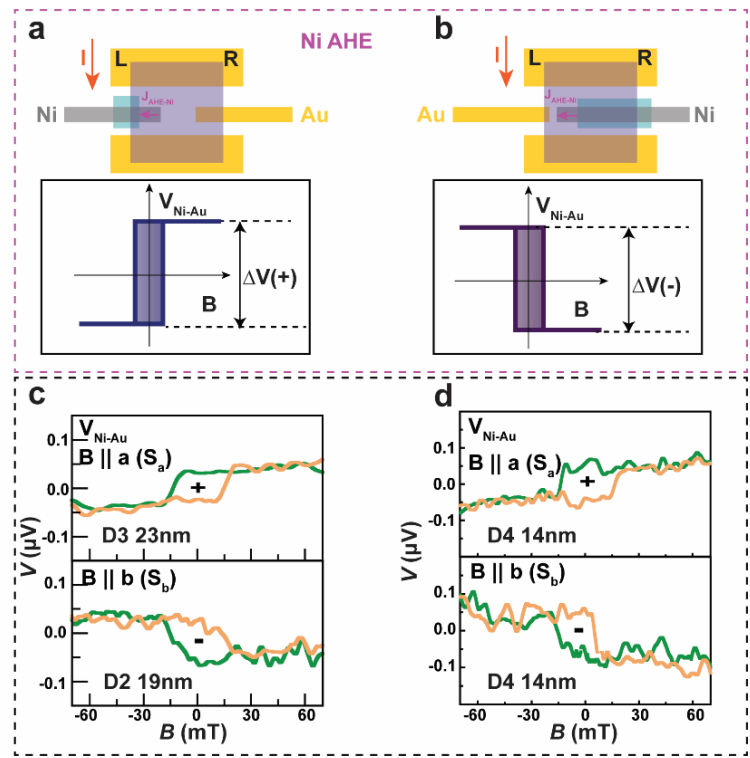

**Fig. S2 | Exclusion of AHE from Ni. a** and **b,** Comparative device schematics for detection at left bulk of sample, but with Ni and Au electrodes extended from opposite directions (top panels). For the same charge current and Ni polarization, current-shunting at the Ni contact will generate anomalous Hall current in the same direction for the two geometries, leading to opposite hysteresis in the field sweep for the Ni-Au voltage measurement (bottom panels). **c** and **d,** Measured hysteresis loops for the two device geometries show the same sign in either $B \parallel a$ or $B \parallel b$ configuration.

## Supplementary Section 3: Bulk CSC Symmetry Analysis

There are two possible CSC mechanisms to generate a uniform $S_a$ distribution on a given surface: the spin Hall Effect (SHE) element $\sigma_{ca}^a$ and the Rashba-Edelstein Effect (REE) element $\alpha_{aa}$. In Fig. 2b, $S_a$ remains relatively constant with the same sign in the left and right bulk, ruling out spin current along the $b$-axis as the origin of this effect.

The REE element $\alpha_{aa}$ induces a uniform bulk magnetization. Conversely, the SHE element $\sigma_{ca}^a$ leads to the generation of opposite spin polarizations $S_a$ at the top and bottom surfaces. To mitigate the effects of MoTe$_2$ degradation in air, our device fabrication process allows for the positioning of the Ni electrode underneath the MoTe$_2$ only. Despite this fixed arrangement, the particular MoTe$_2$ surface in contact with the Ni for each flake is random (depending on the flake orientation with respect to 180° rotation about the $a$-axis), and so through measurements across various samples, we are able to distinguish between these two CSC processes with a high degree of certainty.

As shown in Fig. S3, for REE $\alpha_{aa}$, it is evident that $S_a$ always maintains the same sign regardless of the flake rotation. However, for the SHE element $\sigma_{ca}^a$, $S_a$ will have different signs, yielding both positive and negative hysteresis when measured across different samples by an underlying Ni electrode. As shown in Fig. S4, we only observe $S_a$ hysteresis ($B \parallel a$) of a single sign, allowing us to attribute REE element $\alpha_{aa}$ as dominant mechanism.

Similarly, for $S_b$, the two mechanisms that yield uniform spin polarization on a given surface are the SHE element $\sigma_{ca}^b$ and the REE element $\alpha_{ba}$. As shown in Fig. S3, for REE $\alpha_{ba}$, the resulting $S_b$ displays different sign depending on the flake rotation. In contrast, for SHE element $\sigma_{ca}^b$, $S_b$ always shows the same sign. As shown in Fig. S4, based on the consistent results obtained across multiple samples, we ascertain that $S_b$ stems from the SHE element $\sigma_{ca}^b$.

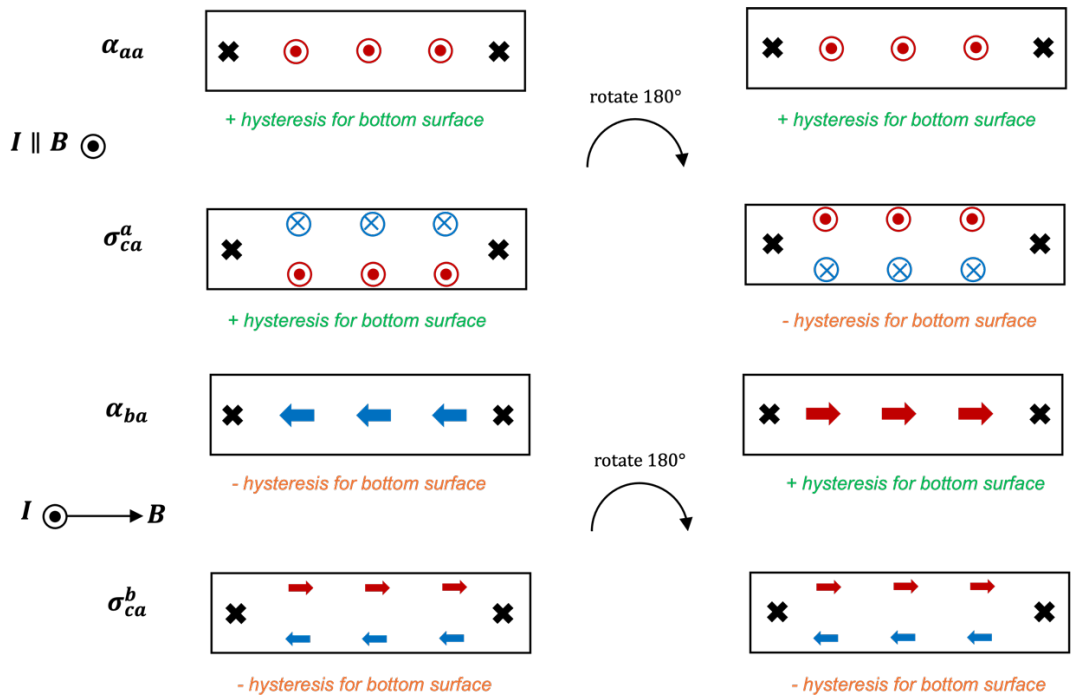

**Fig. S3 | Effect on different bulk CSC mechanisms when rotating sample 180 degrees about the *a*-axis.** The absence of edge spin signals is represented by the thick "x" symbols.

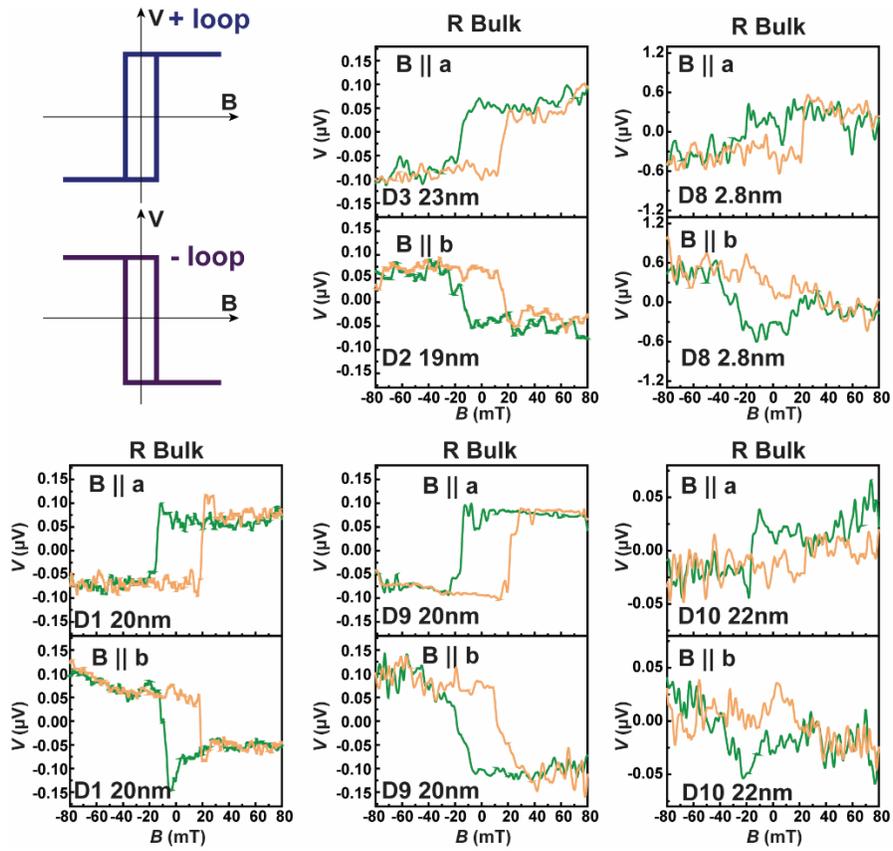

**Fig. S4 | *a*- and *b*-axis spin accumulations measured across different MoTe$_2$ samples.** All *a*- and *b*-axis results show "+" and "-" hysteresis loops, respectively.

## Supplementary Section 4: Effect of sample resistivities on bulk $S_b$ and $S_a$ spin accumulation signal

The spin of bulk $S_b$, which we observe for thicker and ultrathin flakes, are attributed to the SHE element $\sigma_{ca}^b$. This component is allowed and finite according to our calculations for the thick crystal (see Fig. 4b); The expression governing this particular SHE is: $j_c^b = \sigma_{ca}^b E_a$ (for electric field applied the $a$-axis $E_a$, a spin current along the $c$-axis develops with polarization along the $b$-axis). We do not sense the spin current directly but rather measure a voltage signal that is proportional to the net spin density accumulated on the sample surface: $n_\uparrow - n_\downarrow \propto \Delta V \propto t\rho_{cc} j_c^b$. We also current-bias our device, yielding: $E_a = \rho_{aa} j_a$. Here, $\rho_{aa}$ and $\rho_{cc}$ are the resistivities along the $a$- and $c$-axis, respectively, $t$ is the sample thickness, and $j_a$ is the applied charge current density along the $a$-axis. As we normalize the spin accumulation signal to charge current density when comparing across various samples (see Fig. 4a), the relevant experimental quantity is: $\frac{\Delta V}{j_a} \propto t\rho_{aa}\rho_{cc}\sigma_{ca}^b$, which contains several parameters in addition to the spin Hall conductivity.

In Fig. S5, we plot $\rho_{aa}$ measured for different thickness samples, which shows a two-order of magnitude decrease with increasing thickness. This is consistent with results in ref 24. We expect $\rho_{cc}$ to exhibit an even greater change as thinner samples show more significant interlayer shift. Therefore, even if $\sigma_{ca}^b$ increases with increasing thickness as per our calculation in Fig. 4b, the much larger change in sample resistivities will make $\frac{\Delta V}{j_a}$ decrease with increasing thickness.

Similarly, for the REE element $\alpha_{aa}$ giving rise to $S_a$, we expect $n_\uparrow - n_\downarrow \propto \Delta V \propto S_a = \alpha_{aa}\rho_{aa} j_a$, which would yield $\frac{\Delta V}{j_a} \propto \alpha_{aa}\rho_{aa}$. We thus also expect $\frac{\Delta V}{j_a}$ to decrease with increasing thickness even if $\alpha_{aa}$ was thickness-independent, as long as the $a$-axis mirror symmetry is broken.

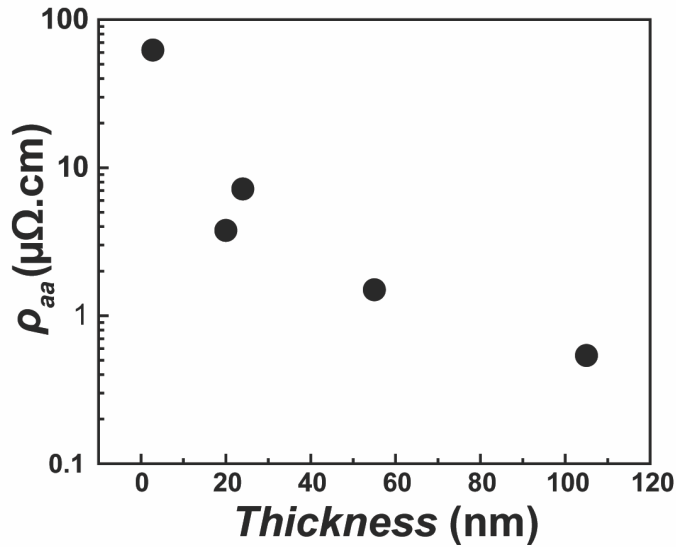

**Fig. S5 | MoTe$_2$ resistivity measured along the $a$-axis vs. sample thickness.**

## Supplementary Section 5: Ab-initio calculation of CSC mechanisms

### Details of crystalline symmetries for the thick and thin samples.

The space group symmetry of the thick sample of $T_d$-MoTe$_2$ is $Pmn2_1$ (No. 31) with the point group $mm2$. This falls under the orthorhombic crystal class. A primitive cell consists of two layers, each layer with 6 atoms (2 Mo and 4 Te). There is no inversion symmetry in the system. There is $M_a$ symmetry, a glide $\overline{M_b}$ symmetry and screw axis $\vec{S_c}$ symmetry along $c-$axis. To model the thin samples, we have considered slab of four layers (4L). This has been achieved by introducing vacuum $> 15$ Å along the $c$-axis. This reduces the space group symmetry of the thin sample of $T_d$-MoTe$_2$ to $Pm$ (No. 6) with the point group $m$. There is no inversion symmetry in the system and has only the $M_a$ symmetry. As we elaborate later, this symmetry has immense consequence in our experimental findings.

### Details of various configurations considered for the thin film samples.

The pristine 4L slab of $T_d$-MoTe$_2$ has $M_a$ symmetry. Keeping this symmetry intact, we have constructed eight configurations by shifting the layers along the $b-$axis in different ways. These configurations have been shown in the main text (Fig. 4) and in Fig. S6. To facilitate easy reference, each configuration is illustrated using three arrows symbolizing the positioning of the top layer relative to the bottom layer. The two pristine configurations that are allowed in the conventional $T_d$-phase represented by →←→ and ←→← and illustrated in the structural schematics of Fig. 4b and Fig. S6a. We considered two configurations shown in Fig. S6e-f where the layer stacking follows →→→ and ←←← patterns. These configurations resemble the $1T'$-phase known in the literature. In addition to these pristine configurations, we considered four mixed configurations. Three of them are displayed in Fig. S6b-d and one highlighted in the main text Fig. 4b. Our *ab-initio* calculations demonstrate that the mixed configurations ←→→ and ←←→ are energetically minimum among the eight configurations as indicated in Fig. S7. The two mixed configurations are illustrated in the structural schematics of Fig. 4b and Fig. S6b.

In addition to the $b$-axis shift, we also investigated $a$-axis shift between the layers in the 4L slab. This configuration has been depicted in Fig. 4b. Unlike the $b$-axis shift which maintains the symmetry of the system, the $a$-axis shift breaks the only remaining $M_a$ symmetry of the systems. To be consistent with recent experimental findings, we considered a scenario where the successive layers are shifted by $l \times 0.1$ Å, where the layer index $l = 1,2,3$.

### Impact of symmetry breaking on the spin Hall conductivities and Rashba-Edelstein coefficients.

The breaking of mirror symmetry $M_a$ plays important role when explaining our experimental findings. The mirror symmetry $M_a$ prohibits the Rashba-Edelstein coefficient $\alpha_{aa}$ ($S_a = \alpha_{aa} E_a$) as well as the $\sigma_{ba}^a$ component of the spin Hall conductivity ($j_b^a = \sigma_{ba}^a E_a$). The explanation is rooted in how the polar and axial vectors transforms under $M_a$. Under mirror operation $M_a$, the pseudovector $S_a$ does not change sign, however the polar vector electric field changes sign, resulting in $\alpha_{aa} = 0$. Similarly, under mirror operation $M_a$, spin current along the $b$ direction with $a$-polarized spin does not change sign, however the sign of the electric field changes, leading to $\sigma_{ba}^a = 0$. In all the configurations where the $M_a$ symmetry is intact, we find that these two coefficients are zero. Only for the configuration with interlayer $a$-axis shift shown in Fig. 4b we find the coefficients to be non-zero.

**Details of the theoretical calculations for the spin Hall conductivities and Rashba-Edelstein coefficients.**

The expression of the spin Hall conductivity implemented in our *ab-initio* calculations is given by

$$\sigma_{ij}^k = e\hbar \int_{BZ} \frac{d\boldsymbol{q}}{(2\pi)^3} \sum_n f_n \Omega_{n,ij}^k(\boldsymbol{q}).$$

Here, $f_n$ is the Fermi-Dirac function and $\Omega_{n,ij}^k(\boldsymbol{q})$ is the spin Berry curvature. The spatial indices $i,j$ and $k$ are the direction of the spin Hall current, applied bias current and spin polarization, respectively. The spin Berry curvature is given by

$$\Omega_{n,ij}^k(\boldsymbol{q}) = -2\text{Im} \sum_{m \neq n} \frac{\langle u_n | \hat{j}_i^k | u_m \rangle \langle u_m | \hat{v}_j | u_n \rangle}{(\epsilon_n - \epsilon_m)^2},$$

where $|u_n\rangle$ is the periodic part of the Bloch wave-function and $\epsilon_n$ is the energy of the *n*-th bands. Here, $\hat{j}_i^k = \frac{1}{2}\{\hat{v}_i, \hat{S}_k\}$ is the spin current operator, $\hat{S}_k$ is the spin operator and $\hat{v}_j = \frac{1}{\hbar}\partial_{q_j} H$ is the velocity operator. To calculate the Rashba-Edelstein coefficients for the 4L slabs, we have implemented the following expression

$$\alpha_{ij} = -g_s \frac{e\tau}{\hbar} \mu_B \int_{BZ} \frac{d\boldsymbol{q}}{(2\pi)^2} \sum_n \frac{\langle u_n | \sigma_i | u_n \rangle}{2} \frac{\partial f_n}{\partial q_j}.$$

Here, $g_s$ is the Lande g-factor, $\tau$ is the scattering time, $\mu_B$ is Bohr magneton and $\sigma_i$ are the Pauli matrices. We have considered $g_s = 2$ and $\tau = 1\,\text{ps}$. For the Brillouin zone integration, we have considered $201 \times 201 \times 201$ k-grid for the thick sample and $201 \times 201$ k-grid for the thin samples. We confirmed the numerical integration convergence for the k-grid.

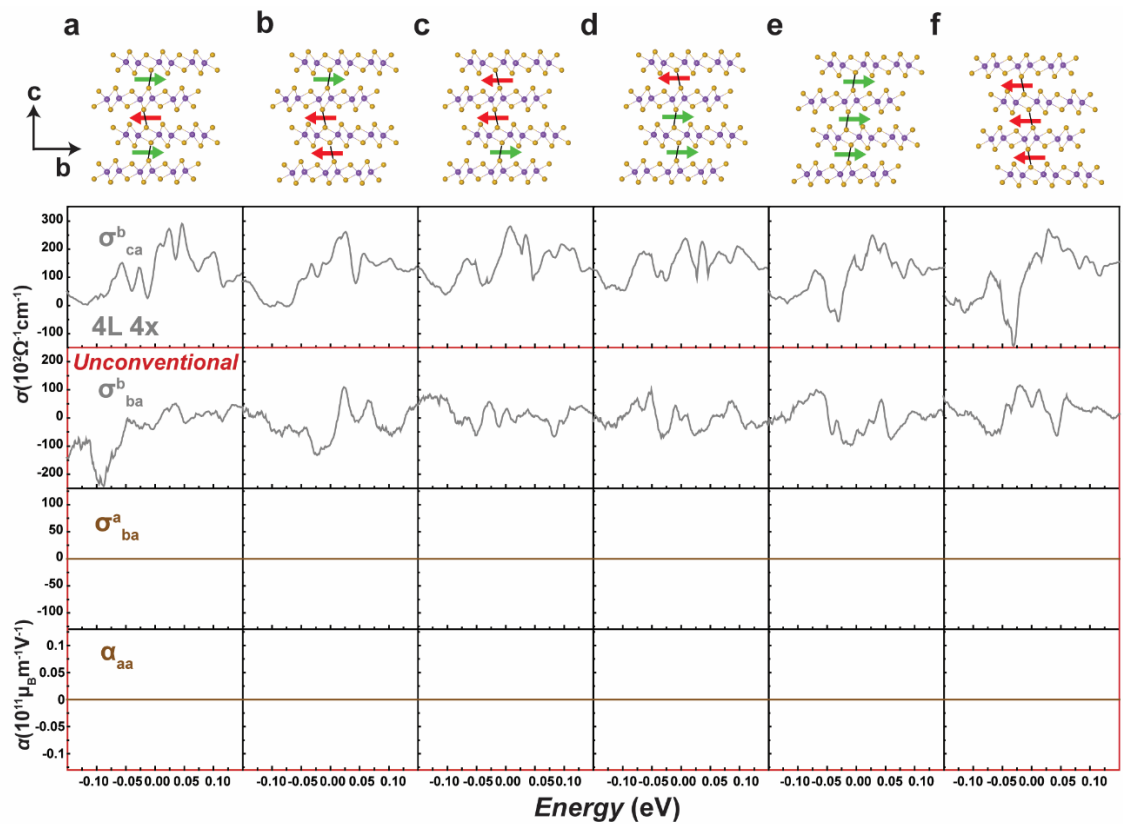

**Fig. S6 | DFT calculations of various CSC mechanisms in four-layer MoTe$_2$. a-f**, The energy variation of CSC mechanisms for each stacking configuration. The corresponding structural schematic of each configuration is shown at the top of each column. The unconventional CSC elements are indicated by the red outlines.

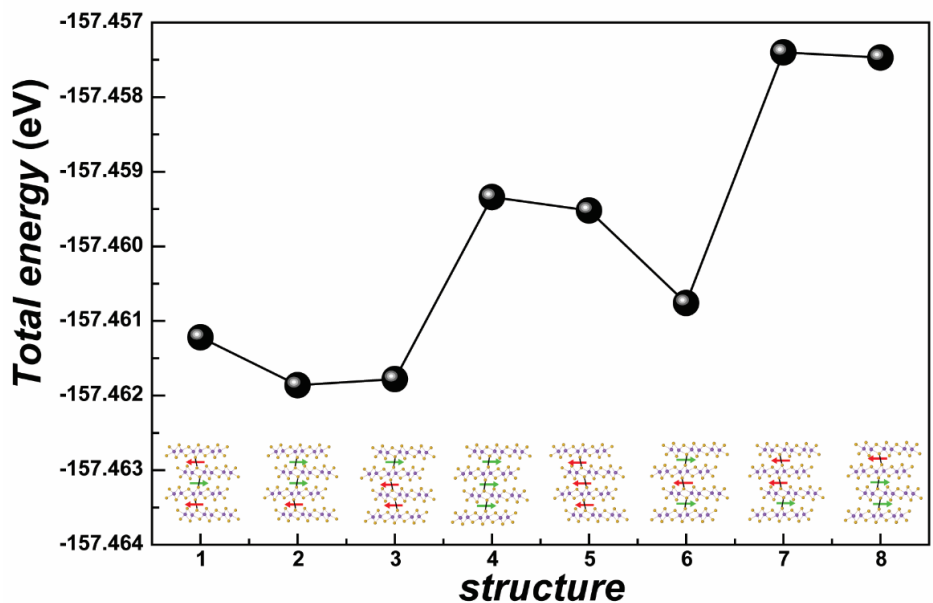

**Fig. S7 | Total energy of the eight stacking configurations of four-layer MoTe$_2$ obtained from DFT.** A structural schematic of each configuration is shown in the inset.

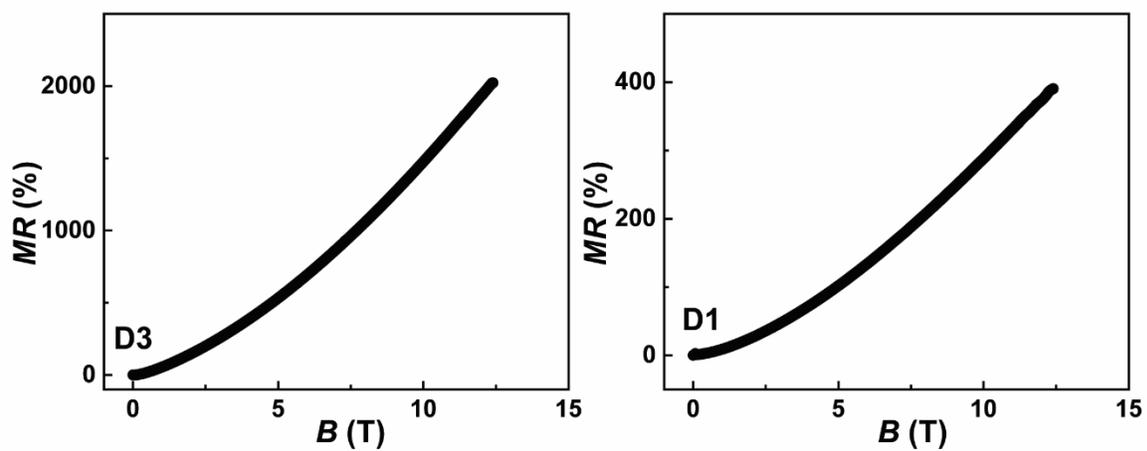

**Fig. S8 | Magnetoresistance behavior.** Percent magnetoresistance as a function of perpendicular magnetic field for devices D3 and D1. The nonlinear trend is consistent with current applied along the MoTe$_2$ a-axis[24].

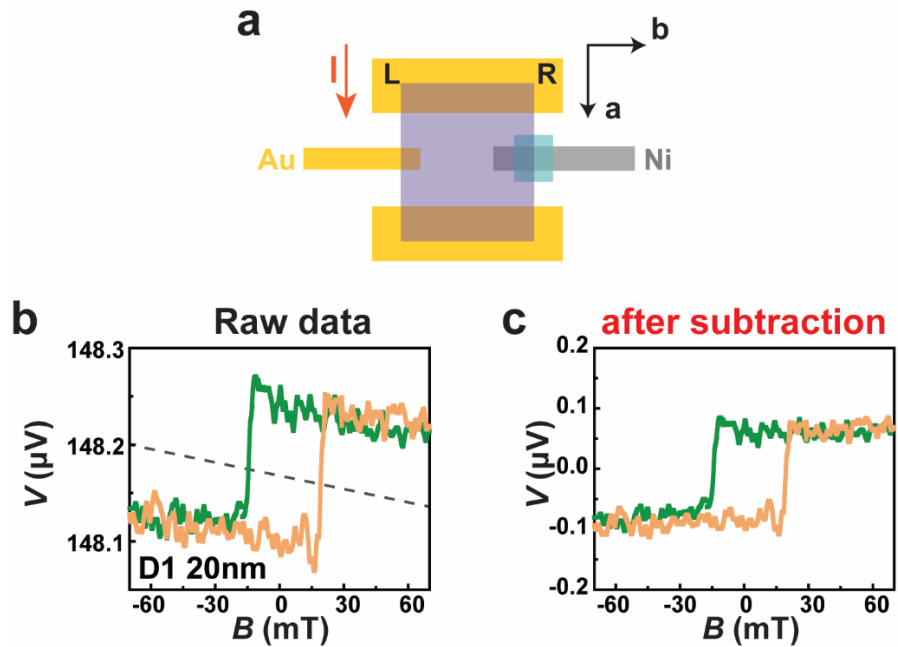

**Fig. S9 | Background subtraction procedure for spin accumulation measurement. a**, Schematic of device. **b**, Voltage difference measured between Ni-Au Hall leads vs in-plane magnetic field for Device 1 (20nm). Upward (downward) field sweep is shown in orange (green). The dashed lines indicate the linear background attributable to the planar Hall effect and/or magnetoresistance. **c**, Same data after background subtraction.

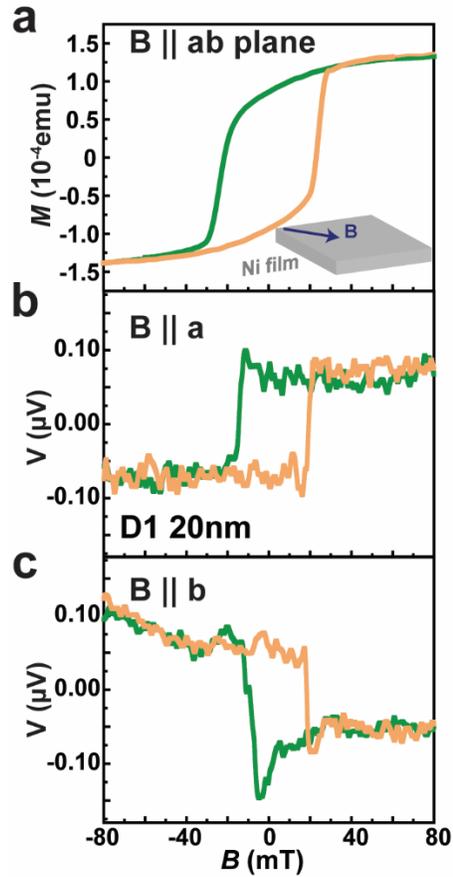

**Fig. S10 | Comparison of Ni magnetization with spin accumulation measurements. a**, Field-dependent magnetization of 35nm-thick Ni film for $B \parallel ab$-plane at 4.4K. **b**, Representative spin accumulation measurements for Device 1 (20nm) at 2K with magnetic field aligned for *a*-axis polarization. **c**, Same for *b*-axis polarization. The coercive fields for the Ni align between the two types of measurements. Upward (downward) field sweep is shown in orange (green) for all data sets.

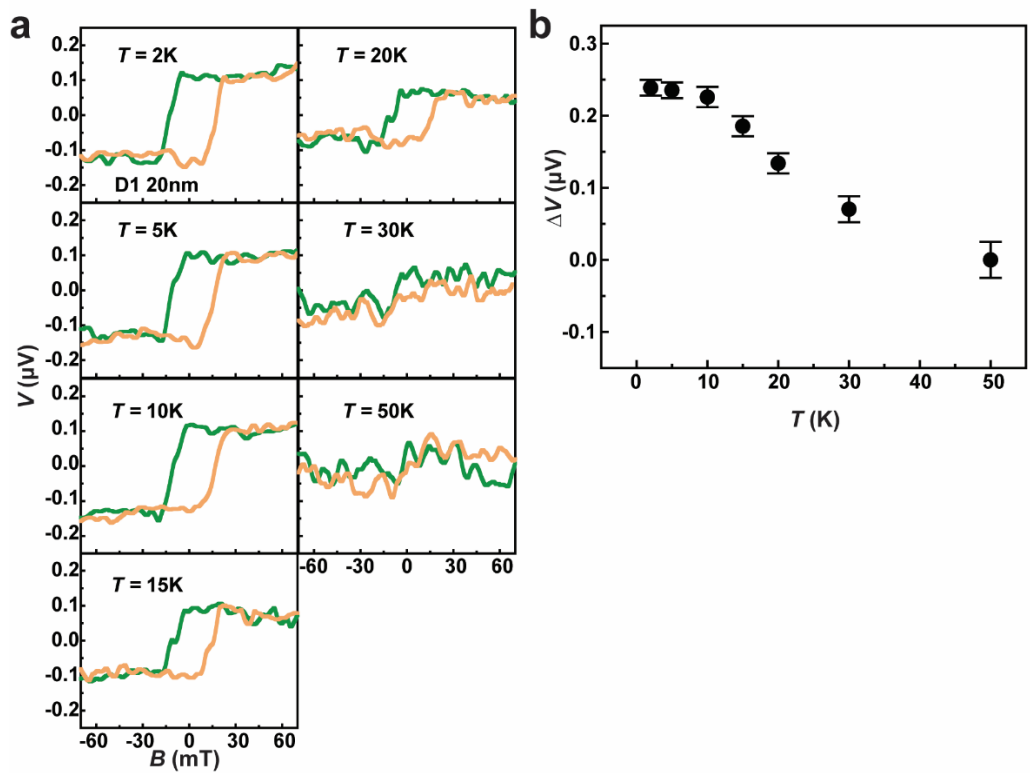

**Fig. S11 | Temperature-dependent spin accumulation measurements. a**, Spin-dependent electrochemical potential $V$ vs. magnetic field ($\theta = -22°$) for bulk Ni contact on Device 1 (20nm) taken at several different temperatures. **b**, Temperature dependence of spin accumulation hysteresis $\Delta V$.

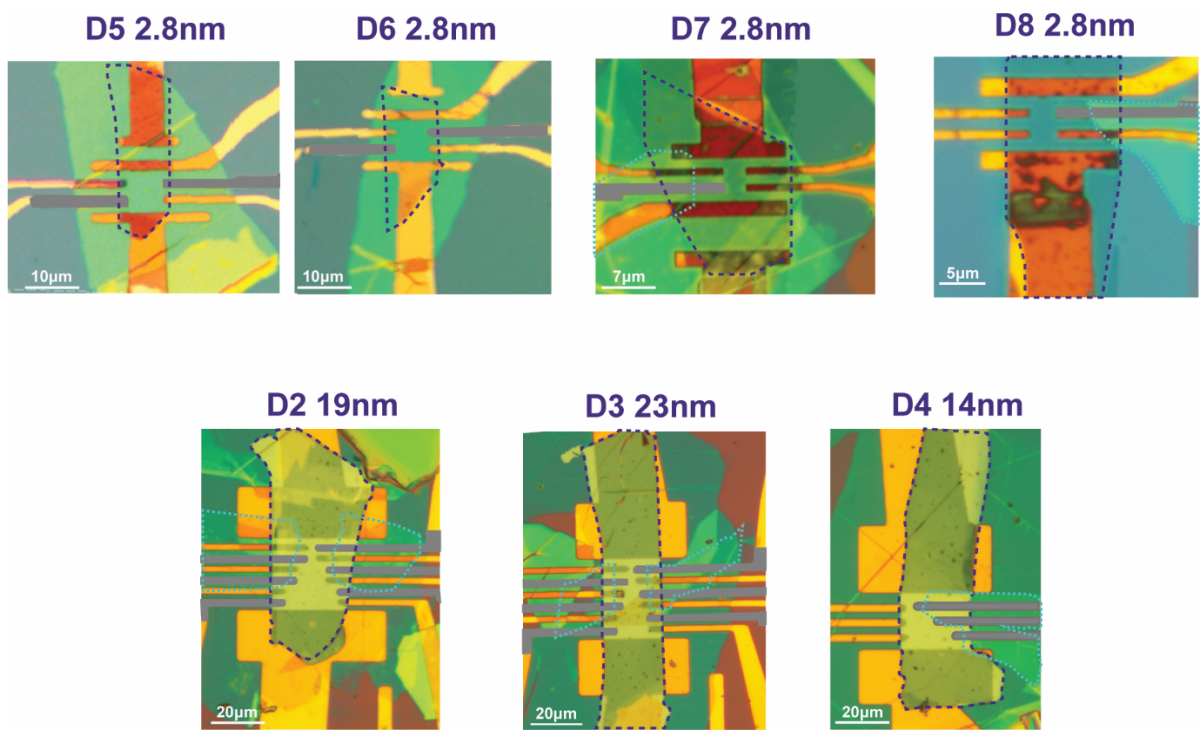

**Fig. S12 | Optical images of MoTe₂ devices with different thicknesses and contact configurations.** MoTe$_2$ and blocking h-BN flakes are outlined in dashed purple and blue, respectively, and the Ni electrodes are colorized in gray for convenience.

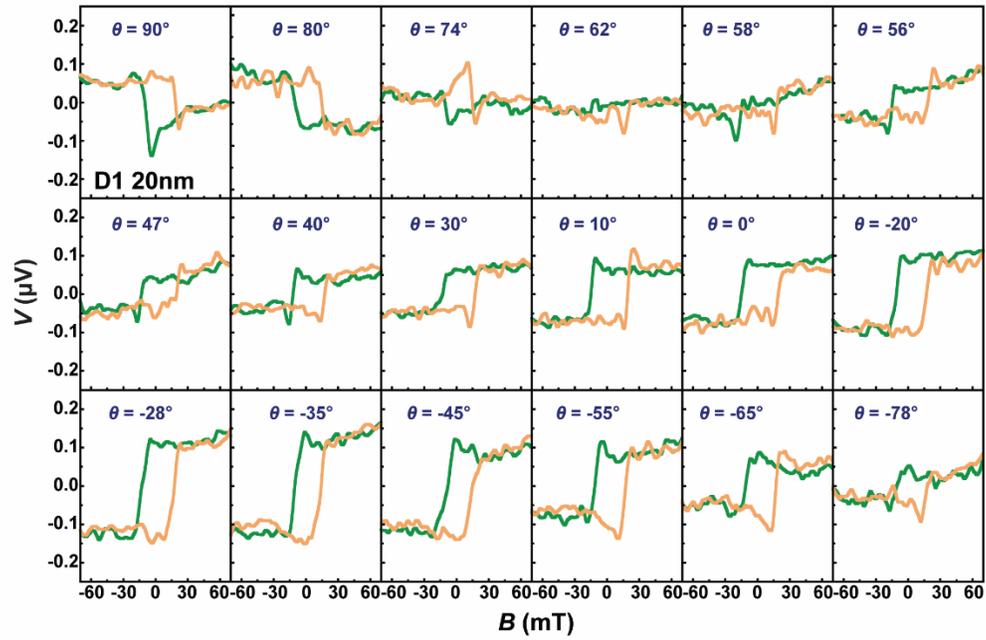

**Fig. S13 | Magnetic-field-angle-dependent spin accumulation measurements.** Spin-dependent electrochemical potential $V$ vs. magnetic field for bulk Ni contact on Device 1 (20nm) taken at various field orientations in the *ab*-plane.

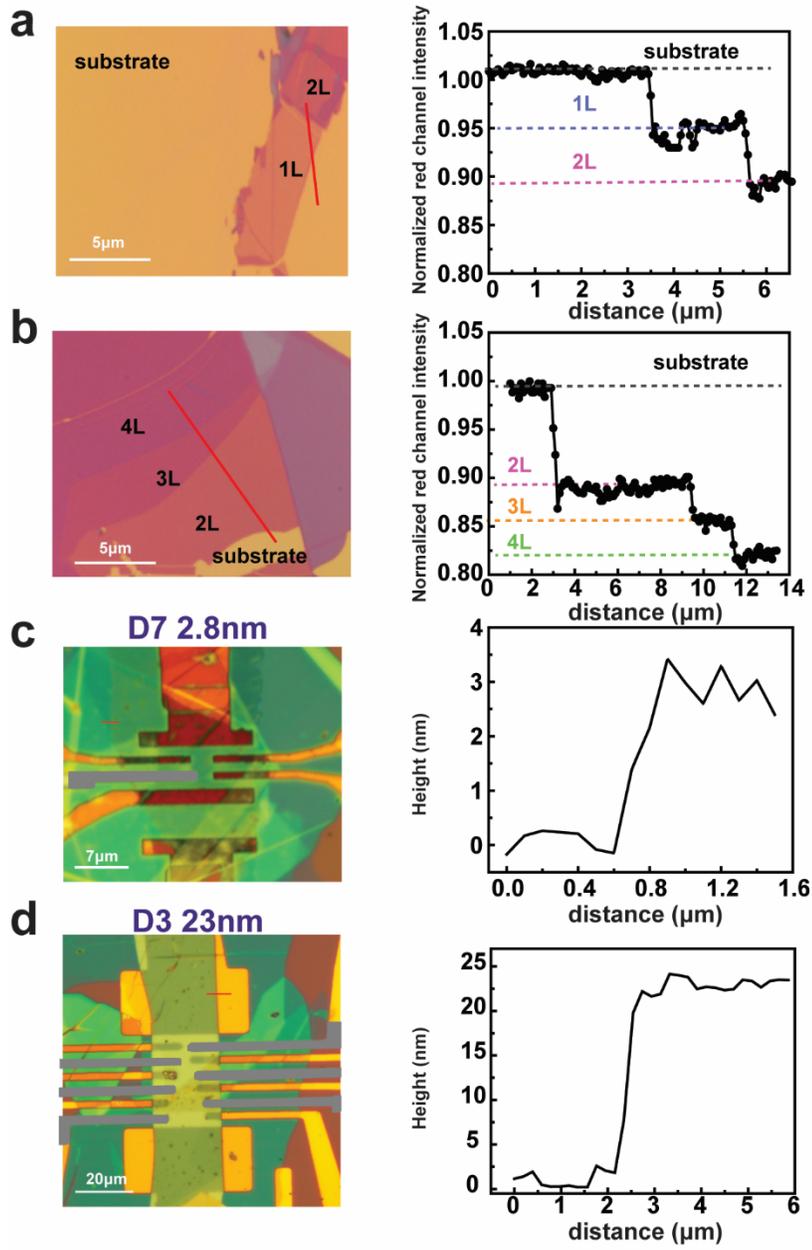

**Fig. S14 | Thickness characterization of MoTe$_2$ flakes/devices. a,b,** Representative optical images and intensity cross-section of MoTe$_2$ thin flakes exfoliated on top of Si substrates with 285nm SiO$_2$ layer. Under consistent illumination conditions, the intensity of the red channel normalized to the substrate can be used to confirm the thickness before device fabrication. **c,d,** Representative height measurements taken using atomic force microscopy on thin and thick MoTe$_2$ devices confirming the sample thickness after fabrication.

**Supplementary Table 1 | List and description of all MoTe₂ devices used in this study.**

| Device # | Thickness | Ni contact position(s) | Field orientation(s) measured |
|---|---|---|---|
| D1 | 20nm | R bulk | B ∥ a,b plane ($\theta = -22°$) |
| D2 | 19nm | L edge, L bulk, R bulk, R edge | B ∥ b |
| D3 | 23nm | L edge, L bulk, R bulk, R edge | B ∥ a |
| D4 | 14nm | L bulk | B ∥ a<br>B ∥ b |
| D5 | 2.8nm | L edge, R edge | B ∥ a |
| D6 | 2.8nm | L edge, R edge | B ∥ b |
| D7 | 2.8nm | L bulk | B ∥ a<br>B ∥ b |
| D8 | 2.8nm | R bulk | B ∥ a<br>B ∥ b |
| D9 | 20nm | R bulk | B ∥ a<br>B ∥ b |
| D10 | 22nm | R bulk | B ∥ a<br>B ∥ b |

**Supplementary Table 2 | Bulk spin axis angle extracted for several devices.** The angle is obtained using $\theta = \arctan(S_b/S_a)$. For D1, this value is slightly different than that obtained from the full angular dependence shown in Fig. 2e.

| Device # | Thickness | $S_a$ (μV/(10⁹A/m²)) | $S_b$ (μV/(10⁹A/m²)) | $\theta$ (degree) |
|---|---|---|---|---|
| D10 | 22nm | 0.044 | -0.035 | -38.5 |
| D1 | 20nm | 0.17 | -0.095 | -29.2 |
| D9 | 20nm | 0.18 | -0.19 | -46.5 |
| D8 | 2.8nm | 0.08 | -0.068 | -40.4 |